\documentclass[12pt]{article}
\pdfoutput=1
\usepackage{amsmath,amssymb,amsfonts,setspace,graphicx,xcolor,cite,upgreek}
\usepackage[utf8]{inputenc}
\usepackage[hidelinks]{hyperref}
\usepackage{textcomp}

\def\O{{\mathcal O}}
\newcommand{\be}{\begin{equation}}
\newcommand{\ee}{\end{equation}}
\newcommand{\comment}[1]{}

\newcommand{\Cint}{C\kern-1em\int}

\def\di{\mathrm{d}}
\def\EP{\textsf{E}\Pcal}
\def\CGHS{$\widehat{\text{CGHS}}$ }
\def\Pcal{\mathcal{P}}
\def\Tr{\text{Tr}}

\def\Pdisk{\Pcal_\textsf{disk}}
\def\Pcyl{\Pcal_\textsf{cylinder}}
\def\multiset#1#2{\ensuremath{\left(\kern-.3em\left(\genfrac{}{}{0pt}{}{#1}{#2}\right)\kern-.3em\right)}}
\usepackage{accents}
\newcommand{\ubar}[1]{\underaccent{\bar}{#1}}

\usepackage{mathtools}

\newtheorem{theorem}{Theorem}
\newtheorem{corollary}{Corollary}

\begin{document}
\begin{center}
{\bf \Large Gravity as an ensemble and the moment problem}
\vskip 1cm
\textbf{Oliver Janssen$^{\star,\dagger}$} and \textbf{Mehrdad Mirbabayi$^\star$} \\
\vskip 0.4cm
{\em $^\star$International Centre for Theoretical Physics \\
$^\dagger$Institute for Fundamental Physics of the Universe \\
Trieste, Italy}
\vskip 0.7cm
including an appendix with \\ {\bf Peter Zograf} \\
\vskip 0.4cm
{\em St. Petersburg Department of the Steklov Mathematical Institute \\ St. Petersburg State University \\ St. Petersburg, Russia}
\end{center}

\vspace{.8cm}

\noindent \textbf{Abstract:} 
If a bulk gravitational path integral can be identified with an average of partition functions over an ensemble of boundary quantum theories, then a corresponding moment problem can be solved. We review existence and uniqueness criteria for the Stieltjes moment problem, which include an infinite set of positivity conditions. The existence criteria are useful to rule out an ensemble interpretation of a theory of gravity, or to indicate incompleteness of the gravitational data. We illustrate this in a particular class of 2D gravities including variants of the CGHS model and JT supergravity. The uniqueness criterium is relevant for an unambiguous determination of quantities such as $\overline{\log Z(\beta)}$ or the quenched free energy. We prove in JT gravity that perturbation theory, both in the coupling which suppresses higher-genus surfaces and in the temperature, fails when the number of boundaries is taken to infinity. Since this asymptotic data is necessary for the uniqueness problem, the question cannot be settled without a nonperturbative completion of the theory.
\vspace{-0.7cm}

\thispagestyle{empty}
\newpage
\setcounter{page}{1}

\section{Introduction}
Long ago Coleman argued that microscopic Euclidean wormholes result in an ensemble of low-energy effective field theories \cite{Coleman}. A variant of this idea could explain the factorization puzzle in the context of holography \cite{Maldacena}: consider a bulk path integral with two identical boundaries, which we denote by $B^2\equiv B \sqcup B$ where $\sqcup$ means disjoint union. Under the {\em conventional} holographic dictionary this must be computing the partition function of two copies of the dual quantum theory living on $B$, i.e. $Z(B)^2$. However, if there are wormhole geometries that connect the two boundaries, the answer differs from the square of the gravity result for $Z(B)$. 

There would be no puzzle if gravitational path integrals were to compute an {\em ensemble average} over boundary duals:
\begin{equation} \label{PisZbar}
	\Pcal(B) \equiv \sum_M \int_{\partial M = B} \mathcal{D}\Phi ~ e^{-S_E} = \overline{Z(B)} \,,
\end{equation}
where $M$ are bulk manifolds and $\Phi$ the collection of all bulk fields. Now it is natural to expect
\be
	\Pcal(B^2) = \overline{Z(B)Z(B)}\neq \Pcal(B)^2 \,.
\ee
Perhaps it is more natural to attribute such puzzles to our poor understanding of gravity, rather than a need to modify the dual quantum description. For instance, it is not clear if additional stringy ingredients in the conventional examples of holography could not cancel the undesired effect of wormholes. 

However, in recent years Jackiw-Teitelboim (JT) gravity \cite{Jackiw,Teitelboim}, a two-dimensional dilaton-gravity model, has emerged as an example of such an ensemble duality. First, as describing a particular low-energy sector of the Sachdev-Ye-Kitaev (SYK) model. SYK is a system of fermions with random couplings \cite{SY,Sachdev,Kitaev}, and several computations in JT gravity coupled to additional matter fields have been shown to agree with disorder averaged quantities in SYK \cite{Maldacena_syk,Kitaev_Suh}. 

Pure JT gravity was later shown to be exactly dual, to all orders in perturbation theory, to a matrix ensemble \cite{Saad}. That is, the JT path integral with $n$ boundaries of lengths $\beta_1,\beta_2,\cdots,\beta_n$ can be thought of as a particular limit, a so-called double scaling (\textsf{DS}) limit, of the following correlation functions in a model of $L\times L$ random Hermitian matrices:
\begin{equation}\label{sss}
	\Pcal_{\textsf{JT}}(\beta_1,\beta_2,\cdots) = \overline{Z(\beta_1)Z(\beta_2)\cdots} = \underset{\textsf{DS}}{\lim} \int \di H ~ e^{-L \, \Tr \, V(H)} \, \text{Tr} \exp \left( - \beta_1 H \right)\text{Tr}\exp \left( - \beta_2 H \right)\cdots \,.
\end{equation}
In the double scaling limit one takes $L \rightarrow \infty$ and focusses on the edge of the spectrum where the density of eigenvalues is controlled by a parameter $e^{S_0}$, which appears in the JT path integral as a coupling constant. The correspondence \eqref{sss} holds then to all orders in $e^{-S_0}$. These results motivated exploring further the possibility of ensemble holography. Indeed several other explicit 2D \cite{Stanford,Iliesiu,Kapec,Maxfield,Witten_W} and 3D \cite{Witten,Hartman} examples have been proposed, as well as hints of such a duality for richer bulk theories \cite{Stanford_noise}.

In this context, an interesting question was recently raised in \cite{Engelhardt}. Assuming an ensemble interpretation of gravity, can we use the knowledge of $\overline{Z(\beta)^n}$ for integer $n$, which we know how to compute in gravity, to extract quantities for which there is no direct gravitational recipe? An important example is the quenched free energy
\be\label{fq}
F_q(\beta) = -\frac{1}{\beta} \, \overline{\log Z(\beta)} \,.
\ee
At low temperature $\beta\to \infty$, $F_q(\beta)$ significantly deviates from its simpler counterpart, the annealed free energy $F_a(\beta) =-(1/\beta) \log\overline{Z(\beta)}$. Hence computing \eqref{fq} is necessary in order to determine low-temperature thermodynamic quantities. In this case the standard approach is the replica method, i.e. analytic continuation in $n$:
\be\label{replica}
\overline{\log Z(\beta)} = \lim_{n\to 0}\frac{1}{n} \left(\overline{Z(\beta)^n}-1\right).
\ee
The authors of \cite{Engelhardt} studied two explicit 2D models of gravity (\CGHS \hspace{-1.2mm}, see \S\ref{CGHSsec}, and JT, see \S\ref{JTsec}) and emphasized that finding the ``correct" analytic continuation is not straightforward, involving perhaps the phenomenon of replica symmetry breaking. Indeed, as noted in \cite{Engelhardt}, without further input the analytic continuation is ambiguous: if $f_\beta(z)$ is an analytic continuation of $\overline{Z(\beta)^n}$ away from the integers, then $F_\beta(z) = f_\beta(z) + g_\beta(z) \sin(\pi z)$ for any (entire) function $g_\beta$ is as well. Moreover computing the limit in \eqref{replica} with $f_\beta$ gives $f_\beta'(0)$ while computing it with $F_\beta$ gives $f_\beta'(0) + \pi g_\beta(0)$. So to compute $\overline{\log Z(\beta)}$ via the replica trick requires picking a preferred analytic continuation. At this stage it is unclear how this should be done, or indeed if it can be.

Our goal here is to point out the relevance of the moment problem to the general discussion of an ensemble interpretation of gravity, and the particular question about averages of non-analytic functions of $Z(B)$ such as the quenched free energy. Suppose we choose the boundary geometry $B$ such that $Z(B) \geq 0$ for a unitary theory. This would for instance be the case if $Z(B)$ were the thermal partition function, in which case we would denote it by $Z(\beta)$ as we have above. If gravity computes an ensemble average over unitary theories, then the sequence
\begin{equation} \label{Pmsequence}
	(\Pcal_n)_{n \geq 0} \equiv \left( 1 , \Pcal(B), \Pcal(B^2), \cdots \right)
\end{equation}
represents the integer moments of a measure $\mu_B$ on the positive real axis:
\be\label{PZdefinitions}
\Pcal(B^n) = \overline{Z(B)^n} \equiv \int_0^\infty \di \mu_B(Z) \, Z^n.
\ee
This implies an infinite set of consistency conditions on this sequence, including
\be
\Pcal(B) = \overline{Z(B)} \geq 0,\qquad \Pcal(B^2) - \Pcal(B)^2= \overline{\left( Z(B) - \overline{Z(B)} \right)^2}\geq 0,\qquad \cdots
\ee
If such a measure exists, it may or may not be unique. For example, $( n+1 )_{n \geq 0}=(1,2,3,\cdots)$ is not the moment sequence of any measure on $\mathbb{R}$ (let alone on $\mathbb{R}^+$), the moments $(n!)_{n \geq 0}$ are unique, among all measures with support on $\mathbb{R}$, to the measure $\di \mu(x) = \di x \, e^{-x}$ which has support on $\mathbb{R}^+$ only, and the measures \cite{Stieltjes1894}
\begin{equation} \label{Stieltjesexample}
	\di \mu(x) = \di x \frac{1}{\sqrt{2 \pi} \, x} \exp \left[ -\frac{1}{2} \left( \log x \right)^2 \right] \left[ 1 + \lambda \, \sin \left( 2 \pi \log x \right) \right] \,, ~~~ \lambda \in [-1,1]
\end{equation}
with support on $\mathbb{R}^+$ all have the same moment sequence $\left( e^{n^2/2} \right)_{n \geq 0}$. Necessary and sufficient conditions on $(\Pcal_n)_{n \geq 0}$ such that a measure exists and is unique are reviewed in \S\ref{momentsec}.

As an application of these conditions, we show in \S\ref{CGHSsec} that $\mu_B$ does not exist for the multi-boundary partition functions of the \CGHS model, a particular 2D dilaton-gravity theory. The same conclusion holds more generally for any gravitational theory in which the connected correlator $\Pcal(B^n)_\textsf{conn}$ vanishes for all but finitely many $n$. Thus either \CGHS gravity is not dual to an ensemble of unitary quantum theories, or its prediction for $\Pcal(B^n)$ is incomplete. This latter possibility is indeed realized in two examples of JT supergravity as we will further discuss in \S\ref{CGHSsec}.

On the other hand if we know that $\mu_B$ exists, whether or not there is an unambiguous answer for $F_q$ (given the gravitational data) depends on the uniqueness of $\mu_B$. This can be inferred from the data $\overline{Z(\beta)^n}$ on the integers alone, i.e. from the moments of $\mu_B$ (see Theorem \ref{BergThill} in \S\ref{momentsec}). The auxiliary machinery of analytic continuation required by the replica trick \eqref{replica} is immaterial: when $\mu_B$ is unique by moments then unambiguously
\begin{equation}
	\overline{\log Z} = \int_0^\infty \di \mu_B(Z) \, \log(Z) \,,
\end{equation}
and when $\mu_B$ is not unique by moments then uniquely determining $\overline{\log Z}$ from the $\overline{Z^n}$ is not possible. In the latter case there does not exist a preferred analytic continuation in \eqref{replica}; the limit is ambiguous and depends on what exactly is meant by ``average" on the left-hand side of \eqref{replica}. In the former case we should note that we have merely proven that the unique determination of $\overline{\log Z}$ (or more generally any $\overline{f(Z)}$) from the $\overline{Z^n}$ is possible \textit{in principle}. We have not given an algorithm which computes $\overline{\log Z}$ from the $\overline{Z^n}$; reconstructing a measure with infinite support from its moments is a difficult task in general (e.g. \cite{shohat1950problem}).

The above motivates the study of the uniqueness problem in the case of JT gravity, where the ensemble dual is known to exist (\S\ref{JTsec}). However, this requires information about the asymptotic behavior of the moments, namely $\overline{Z(\beta)^n}$ as $n \to \infty$. This is not available in the current, perturbative formulation of JT. We will show in the appendix that both the perturbative genus expansion in $e^{-S_0}$, and the low-temperature expansion in $1/\beta$ break down as $n\to \infty$. As a result, without a nonperturbative definition of the theory it is not possible to establish the uniqueness of $\mu_\beta$. This nonperturbative definition (of JT or its dual matrix integral) has in turn been argued in \cite{Saad} to be non-unique.

\section{The Stieltjes moment problem} \label{momentsec}
Among the questions posed in the ``Stieltjes moment problem'' are the following: given a sequence of real numbers $(\Pcal_n)_{n \geq 0}$, when does there exist a (non-negative Borel) measure on the positive real axis for which $\Pcal_k$ is its $k$th moment, and when is this measure unique? These questions were formulated and answered (in some form) by Stieltjes \cite{Stieltjes1894} in 1894, and are still an active research topic today. Among other things one is interested in necessary or sufficient conditions for the existence or uniqueness of a solution (famous sufficient \mbox{(non-)uniqueness} conditions include Carleman's condition \cite{carleman26}, reviewed in Theorem \ref{Carleman} below, and Krein's condition \cite{krein45}), but also in questions such as how to characterize the set of measures with equal moments (see e.g. \cite{BERG199527,krein77,nevanlinna1922asymptotische,pedersen97}). There is a large body of literature on this topic: two classic references are \cite{shohat1950problem,akhiezer1965classical}, a recent book is \cite{Schmudgen17} and a review of checkable\footnote{I.e. those that are formulated directly in terms of a given moment sequence or -- for non-uniqueness results -- a given solution; these are most useful for our problem \eqref{PisZbar}.} \mbox{(non-)uniqueness} criteria is \cite{2017arXiv170301027L}. Extensions of the original Stieltjes moment problem exist where the measure is restricted to have support on a given closed subset of $\mathbb{R}$, as do generalizations to multisequences and multivariate measures with support on a closed subset of $\mathbb{R}^n$. The moment problem on $\mathbb{R}$ is known as the Hamburger moment problem while the moment problem on an interval is known as the Hausdorff moment problem.

Here we simply collect three theorems -- necessary and sufficient conditions for the existence and uniqueness of a solution to a univariate Stieltjes moment problem -- that mathematicians have proven (refs. \cite{Stieltjes1894}, \cite{bergandthill,Berg_Chen_Ismail_2002} and \cite{carleman26} respectively). Some definitions first: we will further abbreviate $\Pcal \equiv ( \Pcal_0,\Pcal_1,\Pcal_2,\cdots )$ and define the shifted sequence $\EP \equiv (\Pcal_1,\Pcal_2,\Pcal_3,\cdots)$. Corresponding to a sequence $\Pcal$ we define for each positive integer $N$ the symmetric $N \times N$ Hankel matrix $H_N(\Pcal)$ where $(H_N)_{ij} = \Pcal_{i-1+j-1}, 1 \leq i,j \leq N$. Finally $\Pcal$ is called positive semidefinite if $H_N(\Pcal)$ is positive semidefinite for all $N$, which we denote by $\Pcal \succeq 0$.

\begin{theorem}[necessary and sufficient for existence] \label{Stieltjestheorem}
	A sequence $\Pcal$ corresponds to the \\moments of a measure on the positive real axis if and only if $\Pcal$ and \textnormal{$\EP$} are positive semidefinite, i.e.
	\begin{equation}
		\Pcal \succeq 0 ~~~ and ~~~ \textnormal{$\EP$} \succeq 0 \,.
	\end{equation}
\end{theorem}
If $\Pcal$ satisfies Theorem \ref{Stieltjestheorem}, it is called a ``Stieltjes moment sequence''. As two examples of these positivity conditions, consider the positivity of $H_2(\Pcal)$ and $H_2(\EP)$. Setting $\Pcal_0 = 1$ the former states $\Pcal_2 - (\Pcal_1)^2 \geq 0$. If $\Pcal_1 = \overline{x}$ and $\Pcal_2 = \overline{x^2}$ are the first and second moments of a (positive) measure then this inequality is satisfied because $\Pcal_2 - (\Pcal_1)^2 = \overline{(x-\overline{x})^2}$, the variance, is the expectation value of a positive random variable. More generally $\overline{p(x)^2}$ must be positive for any polynomial $p$, which is equivalent to the statement that $\Pcal$ is positive semidefinite (e.g. \cite{Schmudgen17}). From $H_2(\EP) \succeq 0$ we get $\Pcal_3 \Pcal_1 - (\Pcal_2)^2 \geq 0$, which must be positive if the measure is supported on the positive real axis because $\Pcal_3 = \overline{x^3}$ and therefore $\Pcal_3 \Pcal_1 - (\Pcal_2)^2 = \overline{x \left( \overline{x}^{1/2} x - \overline{x^2}/\overline{x}^{1/2} \right)^2}$. Requiring that $\overline{x \, q(x)^2}$ is positive for all polynomials $q$ generates the positivity conditions on the $H_N(\EP)$ and vice versa. For the Hamburger moment problem only $\Pcal$ is required to be positive semidefinite: a sequence $\Pcal$ corresponds to the moments of a measure on the whole real axis if and only if $\Pcal$ is positive semidefinite (Hamburger's theorem \cite{Hamburger20}). For the Hausdorff moment problem on $[0,1]$ we require $\Pcal \succeq 0$ and $\EP \succeq \textnormal{\textsf{E}}(\EP)$.
\begin{theorem}[necessary and sufficient for uniqueness] \label{BergThill}
	A Stieltjes moment sequence $\Pcal$ \\corresponds to the moments of exactly one measure on the positive real axis if and only if the smallest eigenvalue of either the $H_N(\Pcal)$ or the $H_N(\textnormal{$\EP$})$ tends to 0 as $N \rightarrow \infty$, i.e.
	\begin{equation}
		\lim_{N \rightarrow \infty} \lambda_{\textnormal{\textsf{min}}}(H_N(\Pcal)) = 0 ~~~ \text{or} ~~~ \lim_{N \rightarrow \infty} \lambda_{\textnormal{\textsf{min}}}(H_N(\textnormal{$\EP$})) = 0 \,.
	\end{equation}
\end{theorem}
Notice that for any sequence $\Pcal$, the sequence $\left( \lambda_\textsf{min}(H_1(\Pcal)),  \lambda_\textsf{min}(H_2(\Pcal)), \cdots \right)$ is decreasing. This can be seen from the identity $\lambda_\textsf{min}(H) = \underset{||c||=1}{\text{min}} \{ H_{ij} c_i c_j \}$ and the fact that $H_N$ is a leading principal submatrix of $H_{N+1}$.

Employing Theorem \ref{BergThill} requires certain control over the smallest eigenvalue of $H_N(\Pcal)$ and $H_N(\EP)$ as $N \rightarrow \infty$. A similar remark pertains to Theorem \ref{Stieltjestheorem}. However Theorem \ref{Stieltjestheorem} could be usefully employed to prove that $\Pcal$ is \textit{not} a moment sequence; we can, perhaps numerically, verify the \mbox{(non-)positivity} of $H_N(\Pcal)$ and $H_N(\textnormal{$\EP$})$ for $N$ as large as is feasible.

It is more stringent for a solution to a Hamburger moment problem to be unique: a Hamburger moment sequence $\Pcal$ corresponds to the moments of exactly one measure on the whole real axis if and only if the smallest eigenvalue of the $H_N(\Pcal)$ tends to 0 as $N \rightarrow \infty$. A solution to the Hausdorff moment problem is always unique.

\begin{theorem}[sufficient for uniqueness] \label{Carleman}
	A Stieltjes moment sequence $\Pcal$ corresponds to the moments of exactly one measure on the positive real axis if
\begin{equation*}
	\sum_{n=1}^\infty \Pcal_n^{-1/2n} = \infty \,.
\end{equation*}
\end{theorem}
Qualitatively, the moments should not diverge too quickly. Notice that this is merely a sufficient uniqueness condition, it is not necessary -- there can be no characterization of uniqueness based on the growth rate of the moments (see \cite{Schmudgen17}). A stronger but more user-friendly result that follows is: 

\begin{corollary}[stronger sufficient uniqueness condition] \label{cor1}
	A Stieltjes moment sequence $\Pcal$ corresponds to the moments of exactly one measure on the positive real axis if there exists a positive constant $c$ such that
\begin{equation} \label{sufficientCarleman}
	\Pcal_n \leq c^n (2n)!
\end{equation}
for all positive integers $n$.
\end{corollary}
For the Hamburger moment problem it is sufficient that the even order moments are bounded, but more stringently so: $\Pcal_{2n} \leq d^n (2n)!$ for a $d > 0$ and all $n$.

\vskip 0.5 cm
\noindent
{\bf Connected components} -- In the context of the ensemble interpretation of gravity, the $\Pcal_n$ above are the full $n$-point correlators $\Pcal(B^n)$. In terms of the connected components we have the following sufficient condition for the uniqueness of the measure on $Z(B)$ (assuming it exists in the first place):
\begin{equation} \label{Pkbound}
	\Pcal_{\textsf{conn},\ell} < c^\ell \, \frac{(2\ell)!}{B_\ell} ~~~ \text{for a } c > 0 \text{ and all } \ell \,.
\end{equation}
Here $B_\ell$ is the $\ell$th Bell number, which counts the number of partitions of a set of size $\ell$ into subsets of any size. For the Hamburger moment problem we have a similar sufficient bound: $\Pcal_{\textsf{conn},\ell} < d^\ell \, \ell!/B_\ell$ for a $d > 0$ and all $\ell$.

To see why \eqref{Pkbound} is sufficient, one can start from the expression of the full $n$-point correlator in terms of the connected components:
\begin{equation} \label{PmPconn}
	\Pcal_n = \sum_{\sum_{\ell=1}^n \ell \, k_\ell = n} (\text{counting factor}) ~~~ \Pcal_{\textsf{conn},1}^{k_1} \Pcal_{\textsf{conn},2}^{k_2} \cdots \Pcal_{\textsf{conn},n}^{k_n} \,,
\end{equation}
where the counting factor counts the amount of ways that a set of $n$ elements can be partitioned into $k_1$ groups of one, $k_2$ groups of two, \dots \phantom{} and $k_n$ groups of $n$.\footnote{A closed-form expression is $n!/ \prod_{i=1}^n (k_i)! (i!)^{k_i}$.} Since the sequence
\begin{equation}
	\left( \frac{(2\ell)!}{B_\ell} \right)^{1/\ell} \,, ~~ \ell = 1,2,\cdots
\end{equation}
is increasing, we deduce from \eqref{Pkbound} that for all $\ell \leq n$
\begin{equation*}
	\Pcal_{\textsf{conn},\ell}^{k_\ell} < c^{\ell k_\ell} \left( \frac{(2\ell)!}{B_\ell} \right)^{k_\ell} \leq \left( c^n \frac{(2n)!}{B_n} \right)^{\ell k_\ell/n} \,.
\end{equation*}
Using this in \eqref{PmPconn} leads to \eqref{sufficientCarleman}.

Eq. \eqref{Pkbound} involves the Bell numbers, which are bounded as follows \cite{Berend10}:
\begin{equation}
B_\ell < \left( 0.792 \, \ell / \log(\ell+1) \right)^\ell	~~~ \text{for all } \ell \in \mathbb{N} \setminus \{0\} \,.
\end{equation}
Using this and $\ell! > \sqrt{2 \pi \ell} \, (\ell/e)^\ell$ we can write a stronger but more familiar-looking sufficient condition,
\begin{equation}
	\exists c > 0 : \forall \ell \in \mathbb{N} \setminus \{ 0 \} : ~~  \Pcal_{\textsf{conn},\ell}^{1/\ell} < c \, \ell \log (\ell+1) \,.
\end{equation}

\vskip 0.5cm
\noindent
{\bf Multivariate measures} -- Finally we mention additional consistency conditions on the ensemble average interpretation of gravity \eqref{PisZbar}. These can be obtained by taking $B = \underset{i=1}{\overset{m}{\bigsqcup}} \, B_i^{n_i}$ in \eqref{PisZbar} and considering the multisequence that arises, which is identified with the moments of a multivariate measure:
\begin{align}
	\Pcal \left( \underset{i=1}{\overset{m}{\bigsqcup}} \, B_i^{n_i} \right) \equiv \Pcal_{n_1 n_2 \cdots n_m} = \int_0^\infty \di \mu_{\sqcup B}(\boldsymbol{Z}) ~ Z_1^{n_1} Z_2^{n_2} \cdots Z_m^{n_m} \,.
\end{align}
This sets up a multidimensional moment problem on $(\mathbb{R}^+)^m$ rather than a univariate one on $\mathbb{R}^+$. Less is known about the solution to this considerably more difficult problem, which is tied to the unknown classification of the non-negative polynomials on $(\mathbb{R}^+)^m$ for $m > 1$ by the Riesz-Haviland theorem \cite{Haviland35}. For example, Stieltjes' existence theorem (Theorem \ref{Stieltjestheorem} above) -- specifically the ``if'' direction -- does not generalize to the $m > 1$ problem. See \cite{Fuglede83,berg87,Schmudgen17} for a discussion of known results, including mention of a necessary existence criterium (also an infinite set of positivity conditions, called ``complete positivity'') and Nussbaum's \cite{nussbaum1965} sufficient existence criterium, and \cite{KLEIBER20137} for a collection of \mbox{(non-)uniqueness} criteria. In this last regard it is worth noting Petersen's theorem \cite{Petersen83}, which states that if $\Pcal_{n_1 n_2 \cdots n_m}$ is a (Stieltjes) moment multisequence which is such that all the marginal (moment) sequences $\Pcal_{n_1 0 0 \cdots 0}, \Pcal_{0 n_2 0 0 \cdots 0}, \cdots, \Pcal_{0 0 \cdots 0 n_m}$ are determinate (i.e. there is a unique univariate measure with these moments), then $\Pcal_{n_1 n_2 \cdots n_m}$ is determinate. The marginals in this case are all of the type $\left( \Pcal(B^n) \right)_{n \geq 0}$ for some choice of $B$. So if we would know that $\Pcal_{n_1 n_2 \cdots n_m}$ is a Stieltjes moment multisequence for all $m \geq 1$, the determinacy of these sequences would follow from the determinacy of the $m = 1$ sequences (for all choices of $B$).

\section{Examples}

\subsection{\CGHS(-like theories)} \label{CGHSsec}
The \CGHS model \cite{Afshar:2019axx} is a simplified version of CGHS dilaton-gravity \cite{Callan:1992rs} involving a metric, two scalars and a $U(1)$ gauge field in two dimensions. The dilaton acts as a Lagrange multiplier in the path integral, selecting flat geometries. In two dimensions these are the disk and the cylinder only. The other scalar is constant on-shell and sets the temperature of black hole solutions. The path integral over {all} two-geometries that have a boundary consisting of $n$ disconnected circles, all of equal length $\beta$, is (from Eq. \eqref{PmPconn}, cf. \cite{Engelhardt})
\begin{equation} \label{PmCGHS}
	\Pcal_n(\beta) = \Pdisk(\beta)^n \sum_{n' = 0}^{\lfloor n/2 \rfloor} \binom{n}{2n'} (2n'-1)!! \, r(\beta)^{n'}  \,,
\end{equation}
where \cite{Afshar:2019tvp,Godet:2020xpk}
\begin{equation}
	r = \frac{\Pcyl}{\Pdisk^2} \,, ~~ \Pdisk = \frac{2\pi}{\beta^2} \,, ~~ \Pcyl = \frac{2 \pi^2}{\beta} \,.
\end{equation}
We have made a choice of units here for the boundary value of the dilaton and for the normalization of the symplectic form in \cite{Afshar:2019tvp,Godet:2020xpk} which corresponds with \cite{Engelhardt}.

Since the only connected contributions to \eqref{PmCGHS} are the disk and the cylinder, we can immediately infer the following Gaussian measure which has the $\Pcal_n(\beta)$ as its moments,
\begin{align}
	\di \mu_\beta(Z) &= \frac{1}{\sqrt{2 \pi} \sigma} \exp \left( \frac{-(Z-\mu)^2}{2 \sigma^2} \right) \di Z \,, \\
	\mu &= \Pdisk \,, ~~~ \sigma^2 = \Pcyl \,.
\end{align}
That this measure is unique follows from Corollary \ref{cor1} in \S\ref{momentsec} (more precisely, its variant for the connected correlators and the Hamburger moment problem stated below \eqref{Pkbound}). Since this measure has support on negative values of $Z$, no identification as an ensemble average over unitary theories can be made for \eqref{PmCGHS}.

It follows from Theorem \ref{Stieltjestheorem} that the sequence $\Pcal = (\Pcal_n)_{n \geq 0}$ is positive semidefinite for all $\beta$, and that $\textsf{E}\Pcal = (\Pcal_n)_{n \geq 1}$ is not positive semidefinite for any $\beta$. For example, one may verify that $H_2(\textsf{E}\Pcal)$ has a negative eigenvalue for all $\beta > \beta_c \equiv 2^{1/3}$ (see \S\ref{momentsec} for the definition of the Hankel matrices), the inverse temperature above which $\Pcyl > \Pdisk^2$ and contributions from wormholes start dominating $\Pcal_n$. For $N \geq 2$, $H_N(\textsf{E}\Pcal)$ is positive semidefinite only on successively smaller intervals $\cdots \subset [0,\beta_{N+1}] \subset [0,\beta_N] \subset \cdots \subset [0,\beta_c]$ with $\lim_{N \to \infty} \beta_N = 0$.

Marcinkiewicz' theorem \cite{Marcink39} states that the Gaussian distributions are the only probability measures on $\mathbb{R}$ with a polynomial cumulant generating function, i.e. with only a finite number of nonzero connected correlators. This implies that gravitational theories which have more than two but only finitely many non-vanishing connected correlators $\Pcal_{\textsf{conn},n}$, cannot be dual to an ensemble. In fact in this case even more general (``nonphysical") ensembles which allow for negative values of $Z$ are not allowed.

The above conclusions hold only if the gravity computation for $\Pcal_n$ is reliable. Otherwise, one could only conclude that the ensemble interpretation implies that the gravitational result is incomplete. For instance, there are two examples of JT supergravity where there is no connected gravitational contribution with more than two boundaries \cite{Stanford}. Hence, on the gravity side these theories fall into the same category as \CGHS \hspace{-1.1mm}. On the other hand, they are shown in \cite{Stanford} to be dual to two Altland-Zirnbauer matrix ensembles with parameters $\upalpha=0,2$ and $\upbeta=2$. In these ensembles $Z(\beta)>0$. Hence it is the nonperturbative contributions to $\Pcal_n$, inferred from the matrix model side, that allow the full answer to be compatible with an ensemble interpretation.

\subsection{JT gravity} \label{JTsec}
As mentioned in the Introduction, for JT gravity Saad, Shenker and Stanford \cite{Saad} showed the existence of an ensemble interpretation \eqref{sss} which implies a measure $\mu_\beta$ on $Z(\beta)$ in \eqref{PZdefinitions}. A remaining question in the context of the moment problem is whether $\mu_\beta$ is unique by moments or not. Theorem \ref{BergThill} in \S\ref{momentsec} gives an answer to this question in principle. Naturally the answer depends on the behavior of $\Pcal_n$ (or $\Pcal_{\textsf{conn},n}$) as $n \rightarrow \infty$. 

This knowledge is not available in the definition of $\Pcal_n$ as an asymptotic perturbative series because it breaks down when $n$ becomes large enough. We prove this statement in detail in the appendix, for two well-known ways in which the JT path integral can be expressed as an asymptotic series, i.e. the genus expansion in $e^{-S_0}$ \cite{Saad} and the $1/\beta$ expansion \cite{Okuyama,Okuyama:2020ncd} that is obtained by a partial resummation of the genus expansion. The structure of the genus expansion and its breakdown are reminiscent of the breakdown of string perturbation theory with large number of external legs \cite{Raju}. On the other hand, the $1/\beta$ expansion has a qualitatively different structure. The leading term at any $n$ coincides with the Airy limit result $\Pcal_n^{\rm Airy}$, which is obtained by taking $S_0\to \infty$ but keeping $x \equiv \beta e^{-2S_0/3}$ fixed.

We conclude that a solely perturbative definition of the JT path integral does not supply us with enough information to determine $\overline{\log Z(\beta)}$ and hence the quenched free energy.\footnote{The need for a nonperturbative completion in order to calculate the quenched free energy has also been emphasized in \cite{Johnson}.} However, it turns out even in the Airy limit where the $\Pcal_n^{\rm Airy}$ are all known (at least in principle) \cite{Okounkov2001}, it is not easy to determine the uniqueness of the distribution. In particular, the moments grow too fast with $n$ to satisfy the sufficient condition for uniqueness in Corollary \ref{cor1}:
\be
\Pcal_n^{\rm Airy}(x) \underset{n\to \infty}\sim \frac{\exp(n^3 x^3/24)}{\sqrt{2\pi} (n x)^{3/2}}.
\ee
In a recent work \cite{Janssen:2021mek} we took a different approach and calculated $\overline{\log Z(\beta)}$ directly in the matrix ensemble, in the limit $\beta \rightarrow \infty$. The result is valid up to perturbative corrections in $1/\beta$ and doubly nonperturbative corrections of order $\exp(-\#_1\exp( \#_2 S_0))$ with $\#_{1,2}= \O(1)$.

\section{Conclusions} \label{conclusions}
The ensemble interpretation of gravity can be tested and refined using the tools that have been developed to study the moment problem. We reviewed some of these tools and applied them to a few simple models of 2D gravity. While these examples were two-dimensional, the connection between the ensemble interpretation and the moment problem holds in any spacetime dimension. As we have seen in the examples of JT gravity and JT supergravity, these tools are often useful in showing when the gravity results are incomplete or ambiguous to determine the properties of the putative ensemble dual. Therefore, they can point to important nonperturbative corrections.

\section*{Acknowledgments}
We thank Christian Berg and Konrad Schm\"udgen for their help in navigating the literature on the moment problem, Sebastian Fischetti for a conversation that lead to this work, and Hamid Afshar, Netta Engelhardt, Victor Godet, Matt Kleban, Alex Maloney and Charles Marteau for discussions. OJ thanks the CCPP at NYU for their hospitality while this work was in progress. The work of PZ was supported by Ministry of Science and Higher Education of the Russian Federation, agreement \textnumero \hspace{0mm} 075-15-2019-1620.

\section*{Appendix: perturbative JT gravity \\ \phantom{} \hfill \normalsize{\textnormal{jointly with Peter Zograf}}}
In this appendix we first review two ways of expressing the JT gravity path integral as an asymptotic series, closely following \cite{Saad,Engelhardt,Okuyama}. These are the genus expansion \eqref{genusexp} and the low-temperature expansion \eqref{betasum}. Along the way we review some properties of the Weil-Petersson volume polynomials that are relevant to JT gravity. Using bounds respected by the intersection numbers of $\psi$-classes on $\overline{\mathcal{M}}_{g,n}$, the Deligne-Mumford compactification of the moduli space of complete hyperbolic surfaces of genus $g$ with $n$ punctures, we show in two subsections that both the genus expansion and the low-temperature expansion fail (in a way that we will clarify) when the amount of boundaries $n$ is taken to infinity.

One way to write the path integral in JT gravity which connects $n \geq 1$ circular boundaries, all of equal length $\beta$, is via the following asymptotic series in the small parameter $e^{-2S_0}$ (the genus expansion):
\begin{equation} \label{genusexp}
	\Pcal_{\textsf{conn},n}(\beta) = \sum_{g=0}^\infty e^{-(2g+n-2) S_0} Z_{g,n}(\beta)
\end{equation}
where
\begin{equation} \label{Zgmbeta}
	Z_{g,n}(\beta) = \left( \prod_{i=1}^n \int_0^\infty \di b_i \, b_i \, Z_\textsf{trumpet}(b_i,\beta) \right) V_{g,n}(\boldsymbol{b})
\end{equation}
for $(g,n) \notin \{ (0,1), (0,2) \}, Z_{0,1}(\beta) = (\sqrt{2 \pi} \beta^{3/2})^{-1} e^{2 \pi^2/\beta}, Z_{0,2}(\beta) = 1/(4 \pi)$ and $Z_\textsf{trumpet}(b,\beta) = (2 \pi \beta)^{-1/2} e^{-b^2/(2\beta)}$ (the same normalization conventions as in \cite{Saad,Engelhardt} are chosen, in which $V_{0,3}(\boldsymbol{b}) = 1$). $V_{g,n}(\boldsymbol{b})$ for all $g,n$ is an even, symmetric polynomial of degree $2(3g+n-3)$ with positive coefficients \cite{mirzakhani07},
\begin{equation} \label{MirzakhaniV0m}
	V_{g,n}(\boldsymbol{b}) = \sum_{|\boldsymbol{\alpha}| \leq 3g+n-3} c_{\boldsymbol{\alpha}}^{(g)} \, b_1^{2 \alpha_1} b_2^{2 \alpha_2} \cdots b_n^{2 \alpha_n} \,,
\end{equation}
where $|\boldsymbol{\alpha}| \equiv \sum_{i=1}^n \alpha_i$. It is the Weil-Petersson volume of the moduli space of genus $g$ Riemann surfaces with $n$ geodesic boundaries of lengths $\boldsymbol{b} = (b_1,b_2,\cdots,b_n)$. Mirzakhani \cite{mirzakhani07} gave a recursion relation that computes the $c_{\boldsymbol{\alpha}}^{(g)}$. They can be written as
\begin{equation} \label{intersectionnos}
	c_{\boldsymbol{\alpha}}^{(g)} = \frac{1}{2^{|\boldsymbol{\alpha}|} \boldsymbol{\alpha}! \, p!} \int_{\overline{\mathcal{M}}_{g,n}} \psi_1^{\alpha_1} \psi_2^{\alpha_2} \cdots \psi_n^{\alpha_n} \omega_1^p
\end{equation}
where $p = 3g+n-3-|\boldsymbol{\alpha}|$, $\boldsymbol{\alpha}! \equiv \alpha_1! \alpha_2! \cdots \alpha_n!$, $\omega_1 = 2 \pi^2 \kappa_1$ where $\kappa_1$ is the first Mumford-Morita-Miller class and the $\psi_j$ are the $\psi$-classes on $\overline{\mathcal{M}}_{g,n}$. Plugging \eqref{MirzakhaniV0m} into \eqref{Zgmbeta} and performing the integrals gives
\begin{align}
	Z_{g,n}(\beta) &= \left( \frac{\beta}{2 \pi} \right)^{n/2} \sum_{|\boldsymbol{\alpha}| \leq 3g+n-3} (2 \beta)^{|\boldsymbol{\alpha}|} \, c_{\boldsymbol{\alpha}}^{(g)} \boldsymbol{\alpha}!  \,.\label{gcontribm}
\end{align}

We have the following useful inequalities \cite{M10,liu11}:
\begin{align}
	c_{\boldsymbol{\alpha}}^{(g)} (2 \boldsymbol{\alpha} + 1)! \, 4^{|\boldsymbol{\alpha}|} &\leq c_{\boldsymbol{0}}^{(g)} = V_{g,n}(\boldsymbol{0}) \,, \label{c0ineq} \\
	V_{g,n}(\boldsymbol{0}) \leq V_{g,n}(\boldsymbol{b}) &\leq e^{|\boldsymbol{b}|/2} V_{g,n}(\boldsymbol{0}) \,, \label{Vgmbineq}
\end{align}
for all $\boldsymbol{\alpha}, g, n$ and $\boldsymbol{b}$. At large $n$ and fixed $g$ we have the result \cite{manin00}
\begin{equation} \label{Vgm0largem}
	V_{g,n}(\boldsymbol{0}) \sim a_g \, C^n n^{(5g-7)/2} n! ~~~ \text{as } n \rightarrow \infty \,,
\end{equation}
for some $a_g, C > 0$, while at large $g$ and fixed $n$ we have \cite{MZ11}
\begin{equation} \label{Vgm0largeg}
	V_{g,n}(\boldsymbol{0}) \sim \frac{\alpha}{\sqrt{g}} (4 \pi^2)^{2g+n-3} (2g+n-3)! ~~~ \text{as } g \rightarrow \infty \,,
\end{equation}
for a $(g,n)$-independent constant $\alpha$ (approximately equal to $1/\sqrt{\pi}$ \cite{Z08}).

We can derive the following two bounds on the terms in the genus expansion \eqref{genusexp} (valid for all $n$ and $g$) from the inequalities \eqref{Vgmbineq}:
\begin{equation} \label{Zgmbounds}
	\left( \sqrt{\frac{\beta}{2 \pi}} \right)^n \leq \frac{Z_{g,n}(\beta)}{V_{g,n}(\boldsymbol{0})} \leq f(\beta)^n \,,
\end{equation}
where
\begin{equation} \label{fofbeta}
	f(\beta) = \int_0^\infty \di b \, b \, Z_\textsf{trumpet}(b,\beta) \, e^{b/2} \,.
\end{equation}

 With the lower bound in \eqref{Zgmbounds} and the asymptotic behavior \eqref{Vgm0largeg} at large $g$ we can demonstrate that the genus expansion \eqref{genusexp} is asymptotic (in the variable $e^{-2S_0}$) for any fixed $n$:
\begin{align}
	e^{-(2g+n-2) S_0} Z_{g,n} &\geq e^{-(2g+n-2) S_0} \left( \frac{\beta}{2 \pi} \right)^{n/2} V_{g,n}(\boldsymbol{0}) \\ &\sim e^{-(2g+n-2) S_0} \left( \frac{\beta}{2 \pi} \right)^{n/2} \frac{\alpha}{\sqrt{g}} (4 \pi^2)^{2g+n-3} (2g+n-3)! \,,
\end{align}
as $g \rightarrow \infty$. It follows that when $g = \mathcal{O}(e^{S_0}) \times \text{(function of $n$)}$ the higher-genus terms in the expansion certainly start dominating.

Using \eqref{gcontribm} in \eqref{genusexp} gives
\begin{align}
	\Pcal_{\textsf{conn},n} &= \left( \frac{x}{2 \pi} \right)^{n/2} (2x)^{n-3} \sum_{\ell = 0}^\infty \beta^{-\ell} \left( \sum_{g=0}^\infty (2x)^{3g} P_{\ell,g,n} \right) \,, \label{betasum} \\
	P_{\ell,g,n} &\equiv \Theta \left( 3g+n-3 - \ell \right) \times 2^{-\ell} \sum_{|\boldsymbol{\alpha}| = 3g+n-3-\ell} c^{(g)}_{\boldsymbol{\alpha}} \boldsymbol{\alpha}! \,, \notag
\end{align}
where $x \equiv \beta \, e^{-2 S_0/3}$ and $\Theta$ is the unit step function. In this expression we have traded the asymptotic genus expansion \eqref{genusexp} in $e^{-2 S_0}$ for an asymptotic expansion in $1/\beta$.\footnote{That this expansion is asymptotic (for fixed $n \geq 3$) can be seen as follows: at any fixed level $\ell \geq n-3$, the coefficient of $\beta^{-\ell}$ will contain a term $(2x)^{3G} 2^{-\ell} c_{\boldsymbol{\alpha}}^{(G)} \boldsymbol{\alpha}!$ where $3G = \ell + 3 - n + q$ with $q = 0,1$ or 2, and $|\boldsymbol{\alpha}| = q$. If we take $\ell \rightarrow \infty$, then $G \rightarrow \infty$. We can then use \cite[Theorem 4.1]{MZ11} to conclude that $c_{\boldsymbol{\alpha}}^{(G)} \sim c_{\boldsymbol{0}}^{(G)}$ which behaves as in \eqref{Vgm0largeg} as $G \rightarrow \infty$. This behavior contains a $(2G)!$ which beats all other factors. The transition happens when $\ell = \mathcal{O}(\beta^{3/2})$.} Note that in contrast to the expression \eqref{genusexp}, the sums over genus in \eqref{betasum} converge, since the partial sums form an increasing sequence which is bounded above:
\begin{align}
	(2x)^{3g} \sum_{|\boldsymbol{\alpha}| = 3g+n-3-\ell} c^{(g)}_{\boldsymbol{\alpha}} \boldsymbol{\alpha}! &\leq \frac{(2x)^{3g}}{4^{3g+n-3-\ell}} V_{g,n}(\boldsymbol{0}) \sum_{|\boldsymbol{\alpha}| = 3g+n-3-\ell} \frac{\boldsymbol{\alpha}!}{(2 \boldsymbol{\alpha}+1)!} \notag \\
	&\leq \frac{(2x)^{3g}}{4^{3g+n-3-\ell}} V_{g,n}(\boldsymbol{0}) \sum_{|\boldsymbol{\alpha}| = 3g+n-3-\ell} (\boldsymbol{\alpha}!)^{-1} \notag \\
	&= \frac{(2x)^{3g}}{4^{3g+n-3-\ell}} V_{g,n}(\boldsymbol{0}) \frac{n^{3g+n-3-\ell}}{(3g+n-3-\ell)!} \notag \\
	&\lesssim h(n) \left[ \pi^{4/3} \frac{2 mx}{3} \right]^{3g} \frac{1}{g!} ~~~ \text{as } g \rightarrow \infty \,, \label{gbound}
\end{align}
where $h$ is an inconsequential function and we've used the bound \eqref{c0ineq} and the asymptotic behavior \eqref{Vgm0largeg}.

In the following two subsections we will investigate the two asymptotic series \eqref{genusexp} and \eqref{betasum} for the same quantities $\Pcal_{\textsf{conn},n}$ (valid in different parameter regimes, however) in the limit where $n$ becomes large. Not surprisingly, we will conclude that the perturbation theory implicit in the asymptotic series fails in both cases. By this we mean the following: at fixed $n$ we can always take $S_0$ or $\beta$ large enough so that the first (few) terms in \eqref{genusexp} and \eqref{betasum} respectively provide a good approximation to the quantity $\Pcal_{\textsf{conn},n}$. Now instead, in view of the bounds discussed in \S\ref{momentsec} related to the moment problem, we would like to know how $\Pcal_{\textsf{conn},n}$ behaves as $n \rightarrow \infty$ while the other parameters are held fixed. Without resumming/nonperturbatively completing \eqref{genusexp} and \eqref{betasum} this knowledge cannot be obtained: as $n \rightarrow \infty$, with $\beta$ and $S_0$ held fixed, the $g = 1$ ($\ell = 1$) term eventually becomes dominant compared to the $g = 0$ ($\ell = 0$) term. We now show this in detail.

\subsection*{Genus expansion fails at large $n$}
In the genus expansion \eqref{genusexp} of $\Pcal_{\textsf{conn},n}$ we consider the ratio of the genus $g+1$ contribution to the genus $g$ contribution, and bound it below as follows:
\begin{align*}
	\frac{e^{-(2(g+1)+n-2)S_0} Z_{g+1,n}}{e^{-(2g+n-2)S_0} Z_{g,n}} &= e^{-2S_0} ~ \frac{\displaystyle\sum_{|\boldsymbol{\alpha}| \leq 3g+n} (2 \beta)^{|\boldsymbol{\alpha}|} \, c_{\boldsymbol{\alpha}}^{(g+1)} \boldsymbol{\alpha}!}{\displaystyle\sum_{|\boldsymbol{\gamma}| \leq 3g+n-3} (2 \beta)^{|\boldsymbol{\gamma}|} \, c_{\boldsymbol{\gamma}}^{(g)} \boldsymbol{\gamma}!} \geq e^{-2S_0} ~ \frac{\displaystyle\sum_{|\boldsymbol{\alpha}| \leq 3g+n-3} (2 \beta)^{|\boldsymbol{\alpha}|} \, c_{\boldsymbol{\alpha}}^{(g+1)} \boldsymbol{\alpha}!}{\displaystyle\sum_{|\boldsymbol{\gamma}| \leq 3g+n-3} (2 \beta)^{|\boldsymbol{\gamma}|} \, c_{\boldsymbol{\gamma}}^{(g)} \boldsymbol{\gamma}!} \,.
\end{align*}
The following inequality holds:
\begin{equation} \label{ZografIneq}
	c_{\boldsymbol{\alpha}}^{(g+1)} > \frac{62 \pi^6}{315} (2g+n-2) c_{\boldsymbol{\alpha}}^{(g)} \,,
\end{equation}
for all $\boldsymbol{\alpha}$ (with $|\boldsymbol{\alpha}| \leq 3g+n-3$), $g$ and $n$. It follows that certainly when $n > \mathcal{O} ( e^{2 S_0} )$ it is no longer justified to neglect any higher-genus contributions.

To derive \eqref{ZografIneq} we start with \cite[(2.12)]{MZ11}, and note
\begin{equation} \label{Bbound}
	(2\boldsymbol{\alpha}+1)! \, 2^{|\boldsymbol{\alpha}|} \,  c_{\boldsymbol{\alpha}}^{(g+1)} \equiv [\tau_{\alpha_1} \tau_{\alpha_2} \cdots \tau_{\alpha_n}]_{g+1,n} > B_{\boldsymbol{\alpha}} > 16 a_3 [\tau_1 \tau_{\alpha_1} \tau_{\alpha_2} \cdots \tau_{\alpha_n}]_{g,n+1}
\end{equation}
where $a_3 = 31 \pi^6/30240$. We will use \cite{DN09,LX09a}
\begin{align}
	(2g+n-2) [\tau_{\alpha_1} \tau_{\alpha_2} \cdots \tau_{\alpha_n}]_{g,n} &= \frac{1}{2} \sum_{l=1}^{3g+n-2} \frac{(-1)^{l-1} l \, \pi^{2l-2}}{(2l+1)!} [\tau_l \tau_{\alpha_1} \tau_{\alpha_2} \cdots \tau_{\alpha_n}]_{g,n+1} \,.
\end{align}
Now, the right-hand side of this equation equals the $l = 1$ term plus a negative contribution. This can be seen from the inequality $[\tau_l \tau_{\alpha_1} \tau_{\alpha_2} \cdots \tau_{\alpha_n}]_{g,n+1} > [\tau_{l+1} \tau_{\alpha_1} \tau_{\alpha_2} \cdots \tau_{\alpha_n}]_{g,n+1}$ (see e.g. \cite[Lemma 3.6]{liu11} or \cite[(4.5)]{MZ11}) and the fact that $\left( l \, \pi^{2l-2}/(2l+1)! \right)_{l \geq 1}$ is a decreasing sequence. Therefore
\begin{equation} \label{mmp1bound}
	(2g+n-2) [\tau_{\alpha_1} \tau_{\alpha_2} \cdots \tau_{\alpha_n}]_{g,n} < \frac{1}{12} [\tau_1 \tau_{\alpha_1} \tau_{\alpha_2} \cdots \tau_{\alpha_n}]_{g,n+1} \,.
\end{equation}
Putting \eqref{Bbound} and \eqref{mmp1bound} together gives \eqref{ZografIneq}.

\subsection*{$1/\beta$ expansion fails at large $n$}
In \eqref{betasum} we would like to compare the $\ell = 1$ contribution to the $\ell = 0$ contribution as $n \rightarrow \infty$ for arbitrary but fixed $x \equiv \beta e^{-2 S_0/3}$, i.e. we are interested in the ratio
\begin{align}
R_n(x) &\equiv \frac{1}{2 \beta} \frac{\sum_{g=0}^\infty (2x)^{3g} \sum_{|\boldsymbol{\alpha}| = 3g+n-4} c_{\boldsymbol{\alpha}}^{(g)} \, \boldsymbol{\alpha}!}{\sum_{g'=0}^\infty (2x)^{3g'} \sum_{|\boldsymbol{\gamma}| = 3g'+n-3} c_{\boldsymbol{\gamma}}^{(g')} \, \boldsymbol{\gamma}!} = \frac{2 \pi^2}{\beta} \frac{\mathcal{F}_1(x,n)}{\mathcal{F}_0(x,n)} \,, \label{Rdef} \\
\mathcal{F}_0(x,n) &= \sum_{g=0}^\infty x^{3g} \sum_{|\boldsymbol{d}| = 3g+n-3} \int_{\overline{\mathcal{M}}_{g,n}} \psi_1^{d_1} \cdots \psi_n^{d_n} \equiv \sum_{g=0}^\infty x^{3g} \mathcal{F}_0^{(g)}(n) \,, \\
\mathcal{F}_1(x,n) &= \sum_{g=0}^\infty x^{3g} \sum_{|\boldsymbol{d}| = 3g+n-4} \int_{\overline{\mathcal{M}}_{g,n}} \psi_1^{d_1} \cdots \psi_n^{d_n} \kappa_1 \notag \\ &= \sum_{g=0}^\infty x^{3g} \sum_{|\boldsymbol{d}| = 3g+n-4} \int_{\overline{\mathcal{M}}_{g,n+1}} \psi_1^{d_1} \cdots \psi_n^{d_n} \psi_{n+1}^2 \equiv \sum_{g=0}^\infty x^{3g} \mathcal{F}_1^{(g)}(n) \,,
\end{align}
where we've assumed $n \geq 4$ and used \eqref{intersectionnos}. From the bound \eqref{gbound} it follows that the series defining $\mathcal{F}_0$ and $\mathcal{F}_1$ have infinite radii of convergence in $x$. (A closed-form expression for $\mathcal{F}_0(x,n)$ is known \cite{Okounkov2001} but we will not use it in the following.) Our strategy is to bound below the individual ratios $\mathcal{F}_1^{(g)}(n) / \mathcal{F}_0^{(g)}(n)$ for each $g$. For $g = 0$ we have the following result due to Kontsevich,
\begin{equation}
	\int_{\overline{\mathcal{M}}_{0,n}} \psi_1^{d_1} \cdots \psi_n^{d_n} = \binom{n-3}{\boldsymbol{d}}
\end{equation}
from which it follows that
\begin{align}
	\mathcal{F}_1^{(0)}(n) &= \frac{(n-2)(n-3)}{2} n^{n-4} \,, \\
	\mathcal{F}_0^{(0)}(n) &= n^{n-3} \,.
\end{align}
So
\begin{equation} \label{genus0}
	\frac{\mathcal{F}_1^{(0)}(n)}{\mathcal{F}_0^{(0)}(n)} = \frac{(n-2)(n-3)}{2n} \,.
\end{equation}
At large $n$ this is bounded below by a linear function in $n$. We aim to show the same is true for all other $g$ too. Using Witten's notation $\langle \tau_{d_1} \tau_{d_2} \cdots \tau_{d_n} \rangle_g \equiv \int_{\overline{\mathcal{M}}_{g,n}} \psi_1^{d_1} \cdots \psi_n^{d_n}$ (where $|\boldsymbol{d}| = 3g+n-3$), the Virasoro constraints imply \cite{Dijkgraaf:1991qh,DIJKGRAAF199159}
\begin{align}
	15 \langle \tau_{d_1} \cdots \tau_{d_n} \tau_2 \rangle_g &= \sum_{j=1}^n \left( 2 d_j + 3 \right) \left( 2 d_j + 1 \right) \langle \tau_{d_1} \cdots \tau_{d_j+1} \cdots \tau_{d_n} \rangle_g + \frac{1}{2} \langle (\tau_0)^2 \tau_{d_1} \cdots \tau_{d_n} \rangle_{g-1} \notag \\ &+ \frac{1}{2} \sum_{\ubar{n} = I \sqcup J} \langle \tau_0 \prod_{i \in I} \tau_{d_i} \rangle_{g'} \langle \tau_0 \prod_{j \in J} \tau_{d_j} \rangle_{g-g'} \,.
\end{align}
Here $\ubar{n}$ is the collection $\{ 1, 2, \cdots, n \}$ which we are to partition into two disjoint subsets $I$ and $J$.

It follows that
\begin{align} \label{F1gnsum}
	15 \mathcal{F}_1^{(g)}(n) > \sum_{|\boldsymbol{d}| = 3g+n-4} \sum_{j=1}^n \left( 2 d_j + 3 \right) \left( 2 d_j + 1 \right) \langle \tau_{d_1} \cdots \tau_{d_j+1} \cdots \tau_{d_n} \rangle \,,
\end{align}
where we have suppressed the index $g$ in the Witten notation because in the following we will only be considering intersection numbers on $\overline{\mathcal{M}}_{g,n}$. We'd like to compare this with
\begin{align}
	\mathcal{F}_0^{(g)}(n) &= \sum_{|\boldsymbol{\alpha}| = 3g+n-3} \langle \tau_{\alpha_1} \cdots \tau_{\alpha_n} \rangle \notag \\
	&= \sum_{\text{0 of } \alpha\text{'s are } 0} \langle \tau_{\alpha_1} \cdots \tau_{\alpha_n} \rangle + \sum_{\text{1 of } \alpha\text{'s is } 0} \langle \tau_{\alpha_1} \cdots \tau_{\alpha_n} \rangle + \cdots + \sum_{\text{$n-1$ of } \alpha\text{'s are } 0} \langle \tau_{\alpha_1} \cdots \tau_{\alpha_n} \rangle \,. \label{F0gsums}
\end{align}
The rewriting in the second equality is possible when $g \geq 1$ (when $g = 0$ at least 3 $\alpha$'s must be zero). Let us consider the first contribution in \eqref{F0gsums}. Every element of this sum can be written in $n$ ways as $\langle \tau_{\alpha_1} \cdots \tau_{\alpha_n} \rangle = \langle \tau_{(\alpha_1-1)+1} \cdots \tau_{\alpha_n} \rangle = \langle \tau_{\alpha_1} \tau_{(\alpha_2-1)+1} \cdots \tau_{\alpha_n} \rangle = \cdots = \langle \tau_{\alpha_1} \tau_{\alpha_2} \cdots \tau_{(\alpha_n-1)+1} \rangle$. Since each of these $n$ appear in the sum in \eqref{F1gnsum} with a coefficient greater than one, we have $\langle \tau_{\alpha_1} \cdots \tau_{\alpha_n} \rangle < (\text{those contributions in \eqref{F1gnsum}})/n$ for every $\langle \tau_{\alpha_1} \cdots \tau_{\alpha_n} \rangle$ appearing in the first sum in \eqref{F0gsums}. Consider then an element in the second sum in \eqref{F0gsums} with, say, $\alpha_1 = 0$. This can be written in $n-1$ ways as $\langle \tau_0 \tau_{\alpha_2} \cdots \tau_{\alpha_n} \rangle = \langle \tau_0 \tau_{(\alpha_2-1)+1} \tau_{\alpha_3} \cdots \tau_{\alpha_n} \rangle = \langle \tau_0 \tau_{\alpha_2} \tau_{(\alpha_3-1)+1} \cdots \tau_{\alpha_n} \rangle = \cdots = \langle \tau_0 \tau_{\alpha_2} \cdots \tau_{(\alpha_n-1)+1} \rangle$, all of which appear in the sum in \eqref{F1gnsum} (and we had not considered previously), namely as
\begin{align}
	(2 \alpha_2+1)(2\alpha_2-1) \langle \tau_0 \tau_{(\alpha_2-1)+1} \tau_{\alpha_3} \cdots \tau_{\alpha_n} \rangle &+ (2 \alpha_3+1)(2\alpha_3-1) \langle \tau_0 \tau_{\alpha_2} \tau_{(\alpha_3-1)+1} \cdots \tau_{\alpha_n} \rangle + \cdots \notag \\
	&+(2 \alpha_n+1)(2\alpha_n-1) \langle \tau_0 \tau_{\alpha_2} \tau_{\alpha_3} \cdots \tau_{(\alpha_n-1)+1} \rangle \\
	&= \langle \tau_0 \tau_{\alpha_2} \cdots \tau_{\alpha_n} \rangle \times \sum_{i=2}^n (2\alpha_i+1)(2\alpha_i-1) \\
	&\geq \langle \tau_0 \tau_{\alpha_2} \cdots \tau_{\alpha_n} \rangle \times \sum_{i=2}^n \left( 4 \alpha_i - 1 \right) \\
	&= \left[ 12(g-1)+3n+1 \right] \langle \tau_0 \tau_{\alpha_2} \cdots \tau_{\alpha_n} \rangle \\
	&> n \langle \tau_0 \tau_{\alpha_2} \cdots \tau_{\alpha_n} \rangle \,,
\end{align}
and therefore, as before, $\langle \tau_0 \tau_{\alpha_2} \cdots \tau_{\alpha_n} \rangle < (\text{the corresponding contributions in \eqref{F1gnsum}})/n$ for every $\langle \tau_0 \tau_{\alpha_2} \cdots \tau_{\alpha_n} \rangle$ in the second sum in \eqref{F0gsums}. Of course the same holds for contributions with any other single $\alpha$ set to zero (and the corresponding contributions in \eqref{F1gnsum} are different from the ones we've considered before). Finally consider the general scenario where $0 \leq m \leq n-1$ $\alpha$'s are set to zero, say, the first $m$. In the same way
\begin{align}
	&(2\alpha_{m+1}+1)(2\alpha_{m+1}-1) \langle (\tau_0)^m \tau_{(\alpha_{m+1}-1)+1} \tau_{\alpha_{m+2}} \cdots \tau_{\alpha_n} \rangle \\ &+ (2\alpha_{m+2}+1)(2\alpha_{m+2}-1) \langle (\tau_0)^m \tau_{\alpha_{m+1}} \tau_{(\alpha_{m+2}-1)+1} \cdots \tau_{\alpha_n} \rangle \notag \\
&+ \cdots + (2\alpha_n+1)(2\alpha_n-1) \langle (\tau_0)^m \tau_{\alpha_{m+1}} \tau_{\alpha_{m+2}} \cdots \tau_{(\alpha_n-1)+1} \rangle \notag \\
&= \langle (\tau_0)^m \tau_{\alpha_{m+1}} \tau_{\alpha_{m+2}} \cdots \tau_{\alpha_n} \rangle \times \sum_{i=m+1}^n (2\alpha_i+1)(2\alpha_i-1) \\
&\geq \langle (\tau_0)^m \tau_{\alpha_{m+1}} \tau_{\alpha_{m+2}} \cdots \tau_{\alpha_n} \rangle \times \sum_{i=m+1}^n (4 \alpha_i - 1) \\
&= \left[ 12(g-1)+3n+m \right] \langle (\tau_0)^m \tau_{\alpha_{m+1}} \tau_{\alpha_{m+2}} \cdots \tau_{\alpha_n} \rangle \\
&> n \langle (\tau_0)^m \tau_{\alpha_{m+1}} \tau_{\alpha_{m+2}} \cdots \tau_{\alpha_n} \rangle \,.
\end{align}
This proves that for $g \geq 1$, $15 \mathcal{F}_1^{(g)}(n) > n \mathcal{F}_0^{(g)}(n)$ for all $n \geq 4$. Including the $g = 0$ result, we have
\begin{equation}
	\frac{\mathcal{F}_1^{(g)}(n)}{\mathcal{F}_0^{(g)}(n)} > \frac{n}{15}
\end{equation}
for all $g \geq 0$ and all $n \geq 5$. This proves that the ratio \eqref{Rdef} diverges as $n \rightarrow \infty$ for any $x > 0$.

\bibliographystyle{utphys}
\bibliography{refs}
\end{document}